\documentclass[10pt]{IEEEtran}

\usepackage[margin=0.73in]{geometry}

\usepackage{hyperref}

\usepackage{times,graphicx,epsfig,amsmath,latexsym,amssymb,verbatim,bbm,amsfonts}
\usepackage{xcolor}

\usepackage{empheq,tikz}

\newcommand{\extra}[1]{}

\usepackage{amsthm}

\newtheorem{theorem}{Theorem}[section]

\newtheorem{corollary}[theorem]{Corollary}

\newtheorem{definition}[theorem]{Definition}
\newtheorem{lemma}[theorem]{Lemma}

\theoremstyle{remark}

\def\squareforqed{\hbox{\rlap{$\sqcap$}$\sqcup$}}
\def\qed{\ifmmode\squareforqed\else{\unskip\nobreak\hfil
\penalty50\hskip1em\null\nobreak\hfil\squareforqed
\parfillskip=0pt\finalhyphendemerits=0\endgraf}\fi}
\def\endenv{\ifmmode\;\else{\unskip\nobreak\hfil
\penalty50\hskip1em\null\nobreak\hfil\;
\parfillskip=0pt\finalhyphendemerits=0\endgraf}\fi}

\newenvironment{proof+}[1]{\noindent \textbf{{Proof #1~} }}{\qed\medskip}

\mathchardef\ordinarycolon\mathcode`\:
\mathcode`\:=\string"8000
\def\vcentcolon{\mathrel{\mathop\ordinarycolon}}
\begingroup \catcode`\:=\active
  \lowercase{\endgroup
  \let :\vcentcolon
  }

\newcommand{\nn}{\nonumber}

\newcommand{\nc}{\newcommand}
\nc{\rnc}{\renewcommand}
\nc{\beq}{\begin{equation}}
\nc{\eeq}{{\end{equation}}}
\nc{\beqa}{\begin{eqnarray}}
\nc{\eeqa}{\end{eqnarray}}
\nc{\lbar}[1]{\overline{#1}}
\nc{\bra}[1]{\langle#1|}
\nc{\ket}[1]{|#1\rangle}
\nc{\ketbra}[2]{|#1\rangle\!\langle#2|}
\nc{\braket}[2]{\langle#1|#2\rangle}
\nc{\proj}[1]{| #1\rangle\!\langle #1 |}
\nc{\avg}[1]{\langle#1\rangle}
\nc{\smfrac}[2]{\mbox{$\frac{#1}{#2}$}}
\nc{\tr}{\operatorname{tr}}
\nc{\tracedist}[1]{\Delta_{}\!\left( #1 \right)}
\nc{\fid}[1]{F\!\left( #1 \right)}

\nc{\ot}{\otimes}
\nc{\dg}{\dagger}
\nc{\dn}{\downarrow}
\nc{\cA}{{\cal A}}
\nc{\cB}{{\cal B}}
\nc{\cC}{{\cal C}}
\nc{\cD}{{\cal D}}
\nc{\cE}{{\mathcal E}}
\nc{\cF}{{\cal F}}
\nc{\cG}{{\cal G}}
\nc{\cH}{{\cal H}}
\nc{\cI}{{\cal I}}
\nc{\cJ}{{\cal J}}
\nc{\cK}{{\cal K}}
\nc{\cL}{{\cal L}}
\nc{\cM}{{\cal M}}
\nc{\cN}{{\cal N}}
\nc{\cO}{{\cal O}}
\nc{\cP}{{\cal P}}
\nc{\cR}{{\cal R}}
\nc{\cS}{{\cal S}}
\nc{\cT}{{\cal T}}
\nc{\cU}{{\cal U}}
\nc{\cX}{{\cal X}}
\nc{\cZ}{{\cal Z}}

\newcommand{\cSbar}{\overline{{\cal S}}}

\nc{\entI}{{\bf I}}
\nc{\entIarrow}{{\bf I}^{\leftarrow}}
\nc{\entH}{{\bf H}}
\nc{\entS}{{\bf S}}
\nc{\entHmin}{\mathbf{H}_{\min}}
\nc{\entHmax}{\mathbf{H}_{\max}}
\nc{\entHtwo}{\mathbf{H}_{2}}

\nc{\entF}{{\bf E}_f}

\nc{\isom}{\simeq}

\nc{\rank}{\operatorname{rank}}
\nc{\ra}{\rightarrow}
\nc{\lra}{\longrightarrow}
\nc{\polylog}{\operatorname{polylog}}
\nc{\poly}{\operatorname{poly}}
\nc{\1}{\mathbb{I}}

\nc{\weight}{\textbf{w}}
\nc{\hamdist}{d_{H}}

\def\d{\delta}
\def\e{\epsilon}

\nc{\Sp}{{{\mathbb S}}}
\nc{\RR}{{{\mathbb R}}}
\nc{\CC}{{{\mathbb C}}}
\nc{\FF}{{{\mathbb F}}}
\nc{\NN}{{{\mathbb N}}}
\nc{\ZZ}{{{\mathbb Z}}}
\nc{\PP}{{{\mathbb P}}}
\nc{\QQ}{{{\mathbb Q}}}
\nc{\UU}{{{\mathbb U}}}
\nc{\OO}{{{\mathbb O}}}
\nc{\EE}{{{\mathbb E}}}
\nc{\id}{{\operatorname{id}}}

\def\ba#1\ea{\begin{align}#1\end{align}}

\nc{\unif}{\textrm{unif}}

\newcommand{\boxfont}[1]{\mathsf{#1}}

\newcommand{\boxP}{\boxfont{P}}
\newcommand{\boxQ}{\boxfont{Q}}
\newcommand{\boxV}{\boxfont{V}}
\newcommand{\boxU}{\boxfont{U}}

\newcommand{\boxPR}{\boxfont{PR}}

\begin{document}

\title{Stronger Attacks on Causality-Based Key Agreement }

\author{\IEEEauthorblockN{Benno Salwey and}
\IEEEauthorblockN{Stefan Wolf}\\
 \IEEEauthorblockA{Faculty of Informatics, Universit\`a della Svizzera Italiana, Via G. Buffi 13, 6900 Lugano, Switzerland}\\
 }

\date{}



\maketitle

\begin{abstract}
Remarkably, it has been shown that in principle, security proofs for quantum key-distribution (QKD) protocols can be independent of assumptions on the devices used and even of the fact that the adversary is limited by quantum theory. All that is required instead is the absence of any hidden information flow between the laboratories, a condition that can be enforced either by shielding or by space-time causality. All known schemes for such Causal Key Distribution (CKD) that offer noise-tolerance (and, hence, must use privacy amplification as a crucial step) require multiple devices carrying out measurements \emph{in parallel} on each end of the protocol, where the number of devices grows with the desired level of security. We investigate the power of the adversary for more practical schemes, where both parties each use a single device carrying out measurements \emph{consecutively}. We provide a novel construction of attacks that is strictly more powerful than the best known attacks and has the potential to decide the question whether such practical CKD schemes are possible in the negative.
\end{abstract}

\section{Introduction}
\begin{figure}[h]
\begin{center}
\includegraphics[width=0.49\textwidth]{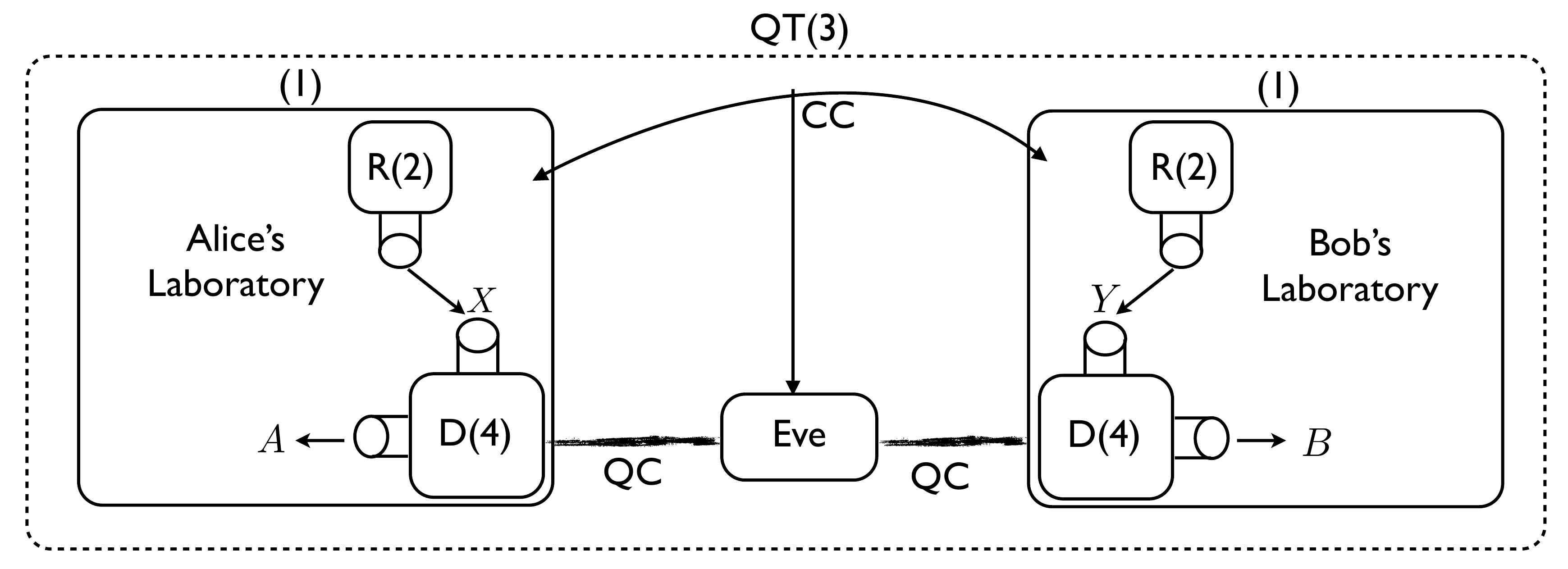}
\end{center}
\caption{\small \sl Schematic setup of QKD protocols with assumptions (1)-(4). The boxes around the legitimate parties'
  laboratories indicate {\bf protection against unwanted information
  leakage\/} (1). The $R$'s are the {\bf sources of free randomness\/}\protect\footnotemark (2) used
  as the inputs $(x,y)$ to the {\bf devices D which generate, and operate on, the specified quantum systems\/} (4). CC refers to a classical insecure (but
  authenticated) channel to which the adversary Eve also has
  access. QC is a completely insecure quantum channel which Eve may
  interfere with to an unspecified extent. The dotted box indicates
  that {\bf the protocol takes place within the rules of quantum theory\/}~(3). }
 \label{fig:DIQKD}
\end{figure}

The use of quantum theory in cryptography allows for realising a task
classically impossible unless assumptions are made  on
the computational power of the adversary: starting from a small
shared secret key, two parties Alice and Bob can generate 
much longer secret keys. Such {\em quantum cryptography\/} goes back
to the celebrated seminal work by Charles Bennett and Gilles Brassard
in 1984~\cite{BB84}. They devised a protocol based on the exchange of
single quantum bits, {\it e.g.\/}, coded into the polarisation of single
photons. The security of the protocol depends on the assumptions sketched in Figure~\ref{fig:DIQKD}.

\begin{figure}[h]
\begin{center}
\includegraphics[width=0.48\textwidth]{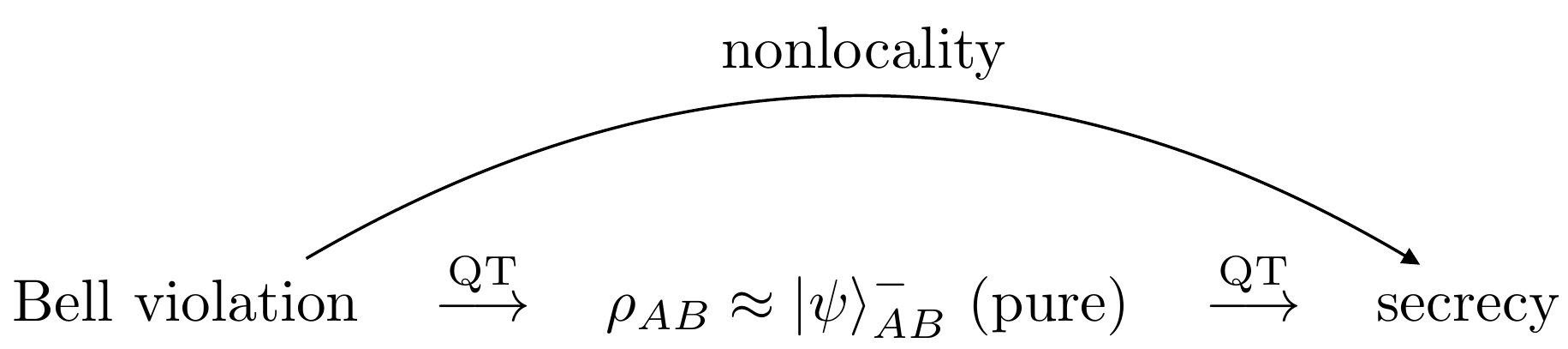}
\end{center}
\caption[LoF entry]{\small \sl Ekert's reasoning: If a system violates the CHSH
  inequality virtually maximally ({\it i.e.\/}, close to Tsirelson's
  bound~\cite{Cirelson1980}), then the framework of quantum theory
  implies that the state of the system must be close to a 
  maximally entangled and, hence, {\bf pure\/} state, a Bell
  state. The purity of the entangled state implies the secrecy of the
  local measurement outcomes. This reasoning is strongly based on the formalism of quantum theory.
  
   Barrett, Hardy, and Kent's reasoning: A
  Bell-inequality violation indicates a non-local correlation that {\bf
    directly\/} implies a constraint on the predictive power of any
  external piece of information (such as, e.g,  Eve's entire
  knowledge) about Alice and Bob's measurement outcomes.
This reasoning is independent of quantum theory.}
\label{fig:EckertsReasoning2}
\end{figure}

\footnotetext{We refer
    to the notion of free randomness used by Colbeck and Renner
    in~\cite{ColRen11noextofQ}: 
    A random variable, generated at some point in space-time, displays free
randomness if it is independent of
any variable which lies outside its future light-cone.
}
It lies in the spirit of cryptography to reduce the assumptions under which
security can be proven.
In the physics community, quantum key distribution became prominent and popular
through the work of Artur Ekert~\cite{Ekert91}, who presented a protocol
based on {\em entangled\/} pairs of quantum bits, and on the phenomenon of
{\em non-local correlations\/}~\cite{Bell64}: If the joint behaviour, under measurements,
of two parts of a system is stronger than what can be explained by shared (classical)
information, one speaks of {\em non-local\/} correlations since no {\em local\/}
hidden-variable model alone can lead to the behaviour (alone). A joint two-partite input-output
behaviour, also called {\em system\/} in the following, is recognised to be non-local
if it violates some Bell inequality, the latter being respected by all local systems.
The rationale of Ekert's method is as follows (see also Figure~\ref{fig:EckertsReasoning2}): If, after exchange and measurement
on the two parts of the entangled pair, respectively, a (virtually) maximal violation of a specific
Bell inequality, due to Clauser, Horne, Shimony, and Holt~\cite{CHSH}, occurs, then
the shared state must be (close to) a maximally entangled pair of quantum bits.
Furthermore, (the completeness of) quantum theory implies that the outcomes when
such a {\em singlet\/} state is measured are (a) perfectly correlated
with each other yet at the same time (b) completely
{\em un\/}correlated with any (classical or quantum) information {\em outside\/} the two laboratories
(and, hence, potentially under an adversary's control); the latter follows from a state
violating maximally the CHSH inequality necessarily being  {\em
  pure}.
Ekert's result (and~\cite{MayYao98} when dealing with noise) has been a big step towards  {\em device-independent\/} security~\cite{ABGMPS07} and the possibility of dropping assumption (4) (see Figure~\ref{fig:DIQKD}). 
Vazirani and Vidick~\cite{VazVid14} devised a scheme similar to Ekert's, where the two parties could each reuse a single device to achieve full device-independent security even tolerating (a certain
level of) noise. They proved that the partial security of the raw key
consisting of the (measurement) outputs of the devices can be
amplified using standard {\em privacy-amplification\/}
techniques~\cite{BBR88PA},~\cite{BBCM95},~\cite{HILL99}. 
However, even their security proof, like Ekert's, rests on the validity of the entire Hilbert-space
formalism of quantum theory. 
It is natural to ask whether it is possible to derive
security of the final key {\em directly and only\/} from the (extent of) non-locality of the generated values (see Figure~\ref{fig:EckertsReasoning2}), together
with the assumption that no hidden communication has taken place between the
laboratories. Barrett, Hardy, and Kent~\cite{BHK2005} have shown
that in principle, the answer is {\em yes\/}: They presented a protocol generating a secret key under the sole assumption that {\em no
illegitimate communication\/}
takes place between the laboratories. Note that such ``{\em causal
key agreement\/}''
requires neither Assumption (3) nor (4) above, see Figure \ref{fig:DIQKD}.

Motivated by this proof of principle, several authors have worked on developing
protocols that are based on the CHSH inequality instead
of the chained Bell inequality~\cite{BC90}, and that are not only more
efficient but also tolerant to
noise~\cite{HRWefficientDIQKD},~\cite{masanesPRL09}. However, besides
the no-signalling assumption {\em between\/} the parties, the
protocols' security proofs must be based on the same condition {\em
  within\/} their laboratories in order to perform privacy
amplification.\footnote{The number of required no-signalling
  conditions is proportional to the negative logarithm of the tolerable noise level.}
Actually, in~\cite{HRWimpossibilityPA}, the impossibility of privacy amplification
was shown if there are no additional no-signalling
conditions assumed. Yet, if Alice and Bob reuse their devices,
then previously obtained {\em outputs cannot depend on future inputs\/} as a
consequence of (2); the corresponding additional conditions are termed
{\em time-ordered no-signalling\/} (TONS)
conditions. In~\cite{RotemLimitsofPA}, it was shown that under the
TONS conditions, super-linear privacy amplification is impossible: 
Using class of attacks which we refer to as ``{\em prefix-code attacks\/}" (see Definition \ref{def:prefixcodeattack}), they showed that if~$n$ is the length of the input to the amplification function, then
the adversary's knowledge on the output is at least of order~$o(1/n)$. Furthermore, prefix-code attacks rule out the use of linear privacy-amplification functions (which are used for $2$-universal hashing) as here the adversary's knowledge on the output remains constant ({\it i.e.\/}, independent of $n$). However, the knowledge prefix-code attacks yield about non-linear functions is limited, {\it e.g.\/}, $\Theta(1/\sqrt{n})$ for majority functions.
We present a novel construction of TONS attacks which comprise prefix-code attacks and, furthermore, can also provide a constant knowledge on the output for highly non-linear functions, {\it i.e.\/}, an improvement of $\Theta(\sqrt{n})$ over prefix-code attacks in the case of majority. That our attack proves TONS privacy amplification with linear functions as well as a highly non-linear function like majority impossible is an indicator that the attack is sufficiently strong to rule out TONS privacy amplification at all. From a practical point of view impossibility of TONS privacy amplification means that Alice and Bob necessarily need additional devices which are shielded against information loss to carry out CKD.

Due to spatial limitations we are forced to omit the detailed proof of Theorem \ref{thm:QStoPReS}, Lemma \ref{lem:majlimit}, and Theorem \ref{thm:SbiasNlarge} and refer the reader to Chapters 3.4.1 and 3.5.4 in~\cite{EveNosignalling}.


\section{Preliminaries}
\label{sec:Preliminaries}
\subsection{No-signalling systems}
We refer to a system $A$ as a black box with an interface consisting of an input $x\in \mathcal{X}$ and an output $a\in \mathcal{A}$, where its complete input-output behaviour is specified by the conditional probability distribution  $\boxP(a\, | \, x)$. If a system $A$ is shared between $m$ parties, each holding $n$ marginal systems, then we denote the
interface of the $i$-th marginal system held by party $j$ by $A^j_i$. 
No-signalling conditions between different systems simply mean that the input one party inserts into her system does not affect the output the other party obtains from her system.
\begin{definition}[$m$-Party no-signalling]
{\rm
An $m$-system box \[\boxP(a^1\dots a^m\, | \, x^1\dots x^m)\] is \emph{$m$-party no-signalling}
if no subset of parties, 
$I^1\subseteq [m]$, can signal to any other (disjoint) subset of parties. Defining $I^2$ to be the complementary set to $I^1$ we have formally
\begin{align}
\label{eq:NS-cond}
 \nn \sum_{a^{I^1}}\boxP(a^{I^1}a^{I^2}\, | \, x^{I^1}x^{I^2})
=\sum_{a_{I^1}}\boxP(a^{I^1}a^{I^2}\, | \, (x')^{I^1}x^{I^2}) \\
\forall I^1, a^{I^2},x^{I^1}, (x')^{I^1},x^{I^2}\ .
\end{align}
We introduce the short-hand notation $A^{I^1}\stackrel{ns}{\lra}A^{I^2}$ if~\eqref{eq:NS-cond} is satisfied, {\it i.e.\/}, the systems $A^{I^1}$ do not signal to the systems~$A^{I_2}$. 
}
\end{definition}
\noindent
\begin{definition}[Marginal]
{\rm
 $A^{I^1}\stackrel{ns}{\lra}A^{I^2}$ induces
a valid {\em marginal\/} distribution $\boxP(a^{I^2}\, | \, x^{I^2})$ on the systems $A^{I^2}$ that is independent of the inputs chosen by the parties in $I^1$.
}
\end{definition}
\noindent
\begin{definition}[No-signalling extension]
{\rm
\label{def:NSextension}
A {\em no-signalling extension\/} of a given system $A$ (possibly consisting of arbitrarily many subsystems), identified with $\boxP(a\, | \, x )$, is any joint system $AE$, identified with $\boxP'(ae\, | \, xu)$, such that $A\stackrel{ns}{\longleftrightarrow}E$ and the marginals on $A$ co\"{\i}ncide, {\it i.e.\/}, $\boxP'(a\, | \, x )=\boxP(a\, | \, x )$.
}
\end{definition}
\noindent
We consider the case of three parties that we identify with Alice, Bob, and Eve
($A^1=A, A^2=B, A^3=E$), where Alice and Bob each hold $n$ subsystems. We use the shorthand notation $A_{\leq n}:=A_1A_2...A_n$ to define the no-signalling conditions that are relevant if Alice and Bob each {\em reuse} their devices to create the systems $A_iB_i$ {\em consecutively}.
\begin{definition}[TONS]
\label{def:TONS}
{\rm
A $(2n+1)$-system \[\boxP(a_{\leq n}b_{\leq n}e\, | \, x_{\leq n}y_{\leq n}u)\] is \emph{time-ordered no-signalling
  (TONS)\/} if no subset of marginal systems can signal to systems outside its causal future. Any union of systems $A_{\leq i}\cup B_{\leq j}\cup E_{\leq k}$, with $k\in\{0,1\}$ and $0\leq i,j\leq n$, must have a valid marginal distribution $\boxP(a_{\leq i}b_{\leq j}e_{\leq k}\, | \, x_{\leq i}y_{\leq j}u_{\leq k})$ induced by the equations
\begin{align}
\label{eq:TONS}
\nn&\sum_{a_{>i}b_{>j}e_{>k}}\boxP(a_{\leq i}a_{>i}b_{\leq j}b_{>j}e_{\leq k}e_{>k}\, | \, x_{\leq i}x_{>i}y_{\leq j}y_{>j}u_{\leq k}u_{>k})\\
\nn&\hspace{4.3cm}= \\
\nn &\sum_{a_{>i}b_{>j}e_{> k}}\boxP(a_{\leq i}a_{>i}b_{\leq j}b_{>j}e_{\leq k}e_{>k}\, | \, x_{\leq i}x'_{>i}y_{\leq j}y'_{>j}u_{\leq k}u'_{>k})\\
\nn& \forall (a_{\leq i},b_{\leq j},e_{\leq k},x_{\leq i},y_{\leq j},u_{\leq k}),(x_{>i},y_{>j},u_{>k}),\\
& (x'_{>i},y'_{>j},u'_{>k}), 0\leq i,j\leq n, k\in\{0,1\} \ .
\end{align}
}
\end{definition}
\noindent
\subsection{Some explicit no-signalling distributions}
\begin{itemize}
\item We denote by $\boxU(a\, | \, x )$ a box that outputs a uniformly random element of the output alphabet ${\cal A}$
\begin{align}
\label{eq:defUNIdist}
\boxU(a\, | \, x )&:=\frac{1}{|{\cal A}|} \qquad \forall a,x \ .
\end{align}
\item We denote by $\boxPR(ab\, | \, xy)$, with ${\cal A},{\cal B},{\cal X},{\cal Y}= \{0,1\}$, as a box with probabilities
\begin{align}
\label{eq:defPRdist}
\boxPR(ab\, | \, xy)&:=\left\{\begin{array}{cc}
\frac{1}{2} & \text{if}\ a\oplus b=x\cdot y\\
0 & \text{otherwise} \ .
\end{array} \right.
\end{align}
\item
We denote by $\boxV(ab\, | \, xy)$, with ${\cal A}=\{0,1\}$ and unspecified alphabets ${\cal B}$, ${\cal X}$, and ${\cal Y}$, as an arbitrary box that satisfies the no-signalling conditions \eqref{eq:NS-cond} and has a uniform marginal on $A$,
\begin{align}
\label{eq:Vbox}
\sum_b\boxV(ab\, | \, xy)=\frac{1}{2}\qquad \forall a,x,y \ .
\end{align}
An example for this type of boxes is the $\boxPR$ box or the boxes corresponding to the chained Bell inequalities~\cite{BC90} considered in~\cite{RotemLimitsofPA} and also multi-partite boxes corresponding to the multipartite \emph{Guess Your Neighbours Input}-game~\cite{GYNI}, since the system $B$ is not specified and can be composed of an arbitrary number of subsystems.
\item We denote by $\boxP_\e(ab\, | \, xy)$ the noisy version of an arbitrary box $\boxP(ab\, | \, xy)$ as the box with probabilities\footnote{We chose this decomposition to be conform with the usual definition of the {\em ``noisy PR-box"} $\boxPR_\e$ when $\boxP$ corresponds to the $\boxPR$ box introduced originally by Popescu and Rohrlich in~\cite{PRbox}.}
 \begin{align}
 \label{eq:noisemix}
 \boxP_\e(ab\, | \, xy):=(1-2\e)\,\boxP(ab\, | \, xy)+2\e\,\boxU(ab\, | \, xy) \ .
 \end{align}
\end{itemize}
\subsection{No-signalling privacy amplification}
 \label{sec:NSattacksPA}
The task of privacy amplification is as follows. Suppose an adversary holding some system $E$ can guess a single bit $a_i$ with probability $1/2+2\e$, but a complete bit-string $a_1 \dots a_n$ only with exponentially small probability, let us say with probability at most $(1/2+2\e)^n$. 
Usually, in a privacy-amplification protocol, one applies a randomly chosen function $f^r$, where $r$ denotes the random choice, to obtain a shorter bit-string $s=f^r(a_1 \dots a_n)$, think of a single bit, that cannot be guessed except with probability (exponentially in $n$) close to $1/2$. 
If the adversary $E$ is governed by classical or quantum theory, it is possible to generate a single bit $s$ that is (exponentially in $n$) close to uniform if the function $f^r$ is chosen uniformly amongst all linear functions~\cite{BBR88PA}, \cite{BBCM95}, \cite{HILL99}, \cite{Ren08}. 
In no-signalling privacy amplification, Alice and Bob hold a box $\boxP(a_{\leq n}b_{\leq n}\, | \, x_{\leq n}y_{\leq n})$, and Alice outputs a Boolean function $f(a_{\leq n})$. To analyse the privacy of such a bit $f(a_{\leq n})$ against a no-signalling adversary, one considers, in analogy to the quantum case, an adversary Eve that holds a ``no-signalling purifying marginal system'' $E$ with input $U$.
\begin{definition}[TONS attack]
{\rm
\label{def:NSadversary}
The box \[\boxP'(a_{\leq n}b_{\leq n}e\, | \, x_{\leq n}y_{\leq n}u)\] is a {\em time-ordered no-signalling (TONS) attack\/} on the box $\boxP(a_{\leq n}b_{\leq n}\, | \, x_{\leq n}y_{\leq n})$ if it is a no-signalling extension of $\boxP(a_{\leq n}b_{\leq n}\, | \, x_{\leq n}y_{\leq n})$ and satisfies the TONS conditions \eqref{eq:TONS}.
}
\end{definition}
\noindent
We study privacy amplification in the context of secret-key distribution. Hence, Alice must communicate her choice $r$ of the privacy-amplification function $f^r(a_{\leq n})$ to Bob eventually, such that they can arrive at a shared secret key in the end of the protocol. Since we assume that Eve can wiretap the classical communication between Alice and Bob and learn the value $r$, she can wait to use her system $E$ until that happens and choose her input as a function of $r$, $u(r)$, accordingly. Her actions are completely specified by the box $\boxP'(a_{\leq n}b_{\leq n}e\, | \, x_{\leq n}y_{\leq n}u(r))$ and the figure of merit is Eve's maximal guessing probability $\boxP'(f^r(a_{\leq n})=e\, | \, x_{\leq n}u(r))$ on the output of the privacy-amplification protocol. 
Since the marginal distribution $\boxP(a_{\leq n}b_{\leq n}\, | \, x_{\leq n}y_{\leq n})$ must be, in particular, independent of $u(r)$, each choice of $r$ can be investigated independently and we can confine our analysis on attacks $\boxP'(a_{\leq n}b_{\leq n}e\, | \, x_{\leq n}y_{\leq n})$ on {\em fixed} functions $f(a_{\leq n})$, where E has no input.
Security against a TONS adversary stems from systems being non-local, {\it i.e.\/}, from systems violating a Bell inequality. If a no-signalling adversary Eve attacks, {\it e.g.\/}, a single $\boxPR_\e(ab \, | \, x y)$ box, the probability $\boxP'(a=e\, | \, x)$ to guess the output $a$ of Alice is at best $1/2+2\e$~\cite{HRWefficientDIQKD}, {\it i.e.}, which is nontrivial exactly if the box is nonlocal.
For simplicity of the representation, we assume that Alice and Bob hold $n$ $\boxV_\e$ boxes, {\it i.e.\/}, the Bell inequality used has binary outcomes on Alice side and we confine ourselves to the hardest case, where Alice outcome is completely random in the noiseless case. The best known previous result on TONS privacy amplification is as follows.
\begin{lemma}~\cite{RotemLimitsofPA}
Assume that Eve attacks $\boxV_\e^{\ot n}(a_{\leq n}b_{\leq n}\, | \, x_{\leq n}y_{\leq n})$ held by Alice and Bob. Then, for any function $f(a_{\leq n})$, there exists a TONS-attack $\boxP'(a_{\leq n}b_{\leq n}e\, | \, x_{\leq n}y_{\leq n})$
 \begin{align}
 \label{eq:Rotemsbias}
 \boxP'(f(a_{\leq n})=e\, | \, x_{\leq n})\geq\frac{1}{2} +\frac{\e}{2n} \qquad \forall x_{\leq n}\ .
 \end{align}
\end{lemma}

\section{The novel attack}
\subsection{Novel construction of TONS attacks}
We present a novel construction of no-signalling attacks on $\boxV_\e^{\ot n}$. 
The idea is to decompose each of the $n$ $\boxV_\e$ boxes in a pure and a noise part via \eqref{eq:noisemix} and then attack each of the $2^n$ terms separately. We identify restrictions \eqref{eq:uniformnessAlice} and \eqref{eq:dontTouchAi} on marginal (classical) distributions $\boxQ_{o-\cS}(a_{\leq n}e)$ on systems $A_{\leq n}E$ that permit extension to a TONS attack for each of the $2^n$ terms in the decomposition of $\boxV_\e^{\ot n}$.
\begin{definition}[Ordered $\cS$-influenceable distributions]
{\rm 
For a set $\cS\in \cP([n])$ we define an \emph{ordered $\cS$-influenceable distribution} $\boxQ_{o-\cS}(a_{\leq n}e)$ as a probability distribution that satisfies uniformity on $a_{\leq n}$
\begin{align}
\label{eq:uniformnessAlice}
\sum_e \boxQ_{o-\cS}(a_{\leq n}e)&=2^{-n}\quad \forall a_{\leq n} \quad \text{and}\\
\label{eq:dontTouchAi}
\boxQ_{o-\cS}(a_i\, | \, a_{<i}e)&=\frac{1}{2}\quad \forall a_{\leq i},e, \quad \text{and} \quad i\in \cSbar \ .
\end{align}
}
\end{definition}
\noindent
We call the distribution $\boxQ_{o-\cS}(a_{\leq n}e)$ ordered $\cS$-influenceable since condition \eqref{eq:dontTouchAi} implies that Eve can only bias the bits $a_i$ with $i\in \cS$, and, furthermore, for $j\notin \cS$ the bits $a_i$ can only be biased with respect to bits $a_j$ if $j<i$.
\begin{definition}[Ordered $(\e,\cS)$-divisible distribution]
\label{def:SgrainESV}
{\rm
Fix a full set of ordered $\mathcal{S}$-influenceable distributions $\{\boxQ_{o-\cS}(a_{\leq n}e)\}$. We define an \emph{ordered $(\e,\cS)$-divisible distribution} $\boxQ_{o-\e}(a_{\leq n}e)$, as
\begin{align}
\label{eq:grainedHSV}
\boxQ_{o-\e}(a_{\leq n}e):=\sum_{\cS\in\cP([n])}\omega(\cS,n,\e)\,\boxQ_{o-\cS}(a_{\leq n}e) \ ,
\end{align}
with weights 
\begin{align}
\label{eq:defweights}
\omega(\cS,n,\e):=(1-2\e)^{n-|\cS|}\,(2\e)^{|\cS|} \ .
\end{align}
}
\end{definition}
\begin{theorem}
\label{thm:QStoPReS}
Any ordered $\cS$-influenceable distribution $\boxQ_{o-\cS}(a_{\leq n}e)$ can be extended to a TONS-attack $\boxP_{o-\cS}(a_{\leq n}b_{\leq n}e\, | \, x_{\leq n}y_{\leq n})$ on the systems $A_{\leq n}B_{\leq n}$ with marginal distribution
\begin{align}
\boxP_{\cS}(a_{\leq n}b_{\leq n}\, | \, x_{\leq n}y_{\leq n}):=\prod_{i\in \cS}\boxU(a_ib_i \, | \, x _iy_i) \prod_{i\in \cSbar}\boxV(a_ib_i \, | \, x _iy_i)
\end{align}
\end{theorem}
\noindent
The proof of Theorem \ref{thm:QStoPReS} consists of an explicit construction of $\boxP'_{\cS}(a_{\leq n}b_{\leq n}e\, | \, x_{\leq n}y_{\leq n})$:
\begin{align}
\label{eq:S-TONSextension1}
\boxP'_{\cS}(e)&=\boxQ_{o-\cS}(e)\\
\label{eq:S-TONSextension2}
\boxP'_{\cS}(a_{\leq n}b_{\leq n}\, | \, x_{\leq n}y_{\leq n}e)&=\prod_{i=1}^n\boxP'_{\cS}(a_ib_i\, | \, a_{<i}b_{<i}x_{\leq n}y_{\leq n}e)\\ 
\label{eq:S-TONSextension3}
\boxP'_{\cS}(a_ib_i\, | \, a_{<i}b_{<i}x_{\leq n}y_{\leq n}e)&=
\left\{\begin{array}{l}
\boxV(a_ib_i \, | \, x _iy_i) \qquad i\in\cSbar \\
\boxU(b_i\, | \, y_i)\,\boxQ_{o-\cS}(a_i\, | \, a_{<i}e)\\
\qquad\qquad\qquad \text{otherwise}\ . 
\end{array} \right.\
\end{align}
It is a bit tedious but straightforward to show that \eqref{eq:S-TONSextension1}-\eqref{eq:S-TONSextension3} implies that $\boxP'_{\cS}(a_{\leq n}b_{\leq n}e\, | \, x_{\leq n}y_{\leq n})$
\begin{enumerate}
\item satisfies the TONS-conditions \eqref{eq:TONS},
\item has the correct marginal on systems $A_{\leq n}B_{\leq n}$:
\begin{align}
\label{eq:SmargAB}
\sum_e \boxP'_{\cS}(a_{\leq n}b_{\leq n}e\, | \, x_{\leq n}y_{\leq n})=\boxP_{\cS}(a_{\leq n}b_{\leq n}\, | \, x_{\leq n}y_{\leq n}) ,
\end{align} 
\item and has the correct marginal on systems $A_{\leq n}E$:
\begin{align}
\label{eq:SmargA}
\sum_{b_{\leq n}} \boxP'_{\cS}(a_{\leq n}b_{\leq n}e\, | \, x_{\leq n}y_{\leq n})=\boxQ_{o-\cS}(a_{\leq n}e) \ .
\end{align} 
\end{enumerate}
\begin{corollary}
\label{thm:ClassictoTONS}
For any ordered $(\e,\cS)$-divisible distribution $\boxQ_{o-\e}(a_{\leq n}e)$, there exists a TONS-attack $\boxP'(a_{\leq n}b_{\leq n}e\, | \, x_{\leq n}y_{\leq n})$ on $\boxV_\e^{\ot n}(a_{\leq n}b_{\leq n}\, | \, x_{\leq n}y_{\leq n})$ such that
\begin{align}
\sum_{b_{\leq n}}\boxP'(a_{\leq n}b_{\leq n}e\, | \, x_{\leq n}y_{\leq n})=\boxQ_{o-\e}(a_{\leq n}e) \qquad \forall x_{\leq n},y_{\leq n}
\end{align}
\end{corollary}
\noindent
Accordingly, we also denote $\boxQ_{o-\e}(a_{\leq n}e)$ as a TONS attack.
\subsection{Prefix-code attacks and their limits}
\begin{definition}[Influence]
\label{def:influence}
{\rm
We define the \emph{influence $\Delta^f(a_{<i})$ of $a_i$ given the prefix $a_{<i}$ on the function $f(a_{\leq n})$} as 
\begin{align}
\label{eq:defInfluence}
\nn\Delta^f(a_{<i})&:=\frac{1}{2}\bigg(\boxQ(f(a_{\leq n})=0\, | \, a_{<i},a_i=0)\\-
&\boxQ(f(a_{\leq n})=0\, | \, a_{<i},a_i=1)\bigg) \ ,
\end{align}
where $\boxQ(a_{\leq n})=2^{-n}$.
}
\end{definition}
\noindent
\begin{definition}[Prefix-code attack]
\label{def:prefixcodeattack}
Given a prefix-code $C=\{c_1, c_2,...,c_k\}$ and the function $f(a_{\leq n})$, we define the corresponding prefix-code attack as the ordered $(\e,\cS)$-divisible distribution $\boxQ_{o-\e}(a_{\leq n}e)$ induced by the set $\{\boxQ_{o-\cS}(a_{\leq n}e)\}$ defined as
\begin{align}
\label{eq:consPREFIXattack1}
\boxQ_{o-\cS}(e)&=\frac{1}{2} \ , \\
\label{eq:consPREFIXattack2}
\boxQ_{o-\cS}(a_i\, | \, a_{<i}e)&=
\left\{\begin{array}{ll}
\frac{1}{2}\left(1+\text{sign}(\Delta^f(c_m))(-1)^{e\oplus a_i}\right)      \\ \quad\text{if}\ \exists m:\ a_{<i}=c_m \cap i\in \cS\ , \\
\frac{1}{2}\qquad \text{otherwise}\ .
\end{array} \right.\
\end{align}
\end{definition}
\noindent
\begin{lemma}
\label{lem:majlimit}
Let the distribution $\boxQ_{o-\e}(a_{\leq n}e)$ be a prefix-code attack on the majority function $\mathrm{Maj}_n(a_{\leq n})$. Then, for any choice of a prefix-code $C=\{c_1,...,c_k\}$ the performance of this attack is
\ba
\nn  \boxQ_{o-\e}(\mathrm{Maj}_n(a_{\leq n})=e)&=\frac{1}{2}+\e\cdot 2^{-n+1}\,\binom{n-1}{\frac{n-1}{2}} \\
n\longrightarrow \infty \qquad &=\frac{1}{2}+\Theta\left(\frac{\e}{\sqrt{n}}\right)\ .
\ea
\end{lemma}
\noindent
The insight behind the proof of Lemma \ref{lem:majlimit} is that in a prefix-code attack $\boxQ_{o-\e}(a_{\leq n}e)$ on $\text{Maj}_n(a_{\leq n})$, a \emph{single} bit $a_i$ is $\e$-biased towards the value $e$, while all other bits $a_{\neq i}$ are uniform when conditioned on $e$; the influence of a single bit $a_i$ on the value of $\text{Maj}_n(a_{\leq n})$ is of the order $\Theta\left(\frac{\e}{\sqrt{n}}\right)$.
\subsection{A stronger attack on Majority}
We construct another attack $\boxQ_{o-\e}(a_{\leq n}e)$ via the set $\{\boxQ_{o-\cS}(a_{\leq n}e)\}$
\begin{align}
\label{eq:majattack1}
\boxQ_{o-\cS}(e)&=\frac{1}{2}\\
\label{eq:majattack2}
\boxQ_{o-\cS}(a_{\leq n}\, | \, e )&=2^{-n+1}\cdot \delta(\text{Maj}_\cS(a_\cS),e)\ ,
\end{align}
for $|\cS|$ being odd (for even $|\cS|$ we define $\text{Maj}_\cS(a_\cS)$ as the majority of all but the last bit).
Intuitively, Eve makes a maximum-likelihood estimate of $\text{Maj}_n(a_{\leq n})$ on the string $a_{\cS}$, which is to compute $\text{Maj}_\cS(a_\cS)$.
Due to the symmetry of the majority function with respect to exchange of indices, the guessing probability $\boxQ_{o-\cS}(\text{Maj}_n(a_{\leq n})=e)$ of the adversary depends only on $s:=|\cS|$.
\begin{theorem}
\label{thm:SbiasNlarge}
Let $|\cS|=s=c\, n$ for some constant $0<c<1$ such that $s$ is odd. Then there exists a series of ordered $\cS$-influenceable distributions $\{\boxQ_{o-\cS}(a_{\leq n}e)\}$ such that
\begin{align}
\label{eq:limitlargeMajbias}
\boxQ_{o-\cS}(\text{Maj}_n(a_{\leq n})=e)\stackrel{n\rightarrow \infty}{=} 1-\frac{\mathrm{arctan}\left(\sqrt{\frac{1-c}{c}}\right)}{\pi}
\end{align}
\end{theorem}
\noindent
Through the concentration of measure around $s=2\e\, n$, induced by the central limit theorem, a direct consequence of Theorem \ref{thm:SbiasNlarge} is Corollary \ref{cor:Majiskilled}.
\begin{corollary}
\label{cor:Majiskilled}
For any $\delta>0$, there exists a series of $\boxQ_{o-\e}(a_{\leq n}e)$ such that
\begin{align}
\label{eq:Majiskilled}
\boxQ_{o-\e}(Maj_n(a_{\leq n})=e)\stackrel{n\rightarrow \infty}{\geq}&1-\frac{\mathrm{arctan}\left(\sqrt{\frac{1-(2\e-\d)}{(2\e-\d)}}\right)}{\pi} \ .
\end{align}
\end{corollary}
\noindent
Lemma \ref{lem:majlimit} and Corollary \ref{cor:Majiskilled} imply an $\Theta(\sqrt{n})$ advantage of our attack on the best previously known attack, the prefix-code attack.
\section{Conclusion}
Causal key distribution (CKD) requires only a {\em minimal} set of assumptions, {\it i.e.}, (1) a shielded laboratory and (2) free randomness, see Figure \ref{fig:DIQKD}, which both can be considered also {\em necessary}: If the parties' laboratories leak information about the key the adversary eventually learns it. Without free randomness everything becomes deterministic from the view of the adversary, and she can compute the key herself.
All CKD protocols that offer noise tolerance~\cite{HRWefficientDIQKD},~\cite{masanesPRL09} have the impractical requirement for Alice and Bob to use many devices in parallel, where each device needs to be shielded against unwanted information leakage individually. We address the (still) open problem whether CKD is also possible if Alice and Bob each {\em reuse a single device} and construct a novel attack on the necessary time-ordered no-signalling (TONS) privacy-amplification step in the CKD protocol. Our construction is a generalisation of the best known attack~\cite{RotemLimitsofPA}, and we prove it to be superior if majority functions are used for TONS privacy amplification; the amount of knowledge that our attack provides is optimal (up to a constant factor). That our attack performs well against TONS privacy amplification with linear functions as well as with a highly non-linear function like majority may suggest that it also powerful enough to prove impossibility of TONS privacy amplification in general, {\em if} this is indeed the case.

\section*{Acknowledgments}
The authors thank Rotem Arnon-Friedman, \"{A}min Baumeler, Gilles Brassard, Omar Fawzi, Arne Hansen, Karol Horodecki, Jibran Rashid, Renato Renner, and Dave Touchette for
stimulating discussions and helpful comments. BS and SW are supported 
by the Swiss National Science Foundation (SNF), the NCCR {\em QSIT}, by the COST action on
``Fundamental Problems in Quantum Theory,'' and the CHIST-ERA project
{\em DIQIP}. 
 
 \bibliographystyle{plain}
\bibliography{NonLocality}

\begin{thebibliography}{10}

\bibitem{ABGMPS07}
Antonio Ac\'in, Nicolas Brunner, Nicolas Gisin, Serge Massar, Stefano Pironio,
  and Valerio Scarani.
\newblock Device-independent security of quantum cryptography against
  collective attacks.
\newblock {\em Phys. Rev. Lett.}, 98:230501, Jun 2007.

\bibitem{GYNI}
Mafalda~L. Almeida, Jean-Daniel Bancal, Nicolas Brunner, Antonio Ac\'in,
  Nicolas Gisin, and Stefano Pironio.
\newblock Guess your neighbor's input: A multipartite nonlocal game with no
  quantum advantage.
\newblock {\em Phys. Rev. Lett.}, 104:230404, Jun 2010.

\bibitem{RotemLimitsofPA}
Rotem Arnon-Friedman and Amnon Ta-Shma.
\newblock Limits of privacy amplification against nonsignaling memory attacks.
\newblock {\em Phys. Rev. A}, 86:062333, Dec 2012.

\bibitem{BHK2005}
Jonathan Barrett, Lucien Hardy, and Adrian Kent.
\newblock No signaling and quantum key distribution.
\newblock {\em Phys. Rev. Lett.}, 95:010503, Jun 2005.

\bibitem{Bell64}
John~S. Bell.
\newblock On the {Einstein-Podolsky-Rosen} paradox.
\newblock {\em Physics}, 1:195--200, 1964.

\bibitem{BB84}
Charles~H. Bennett and Gilles Brassard.
\newblock Quantum cryptography: Public key distribution and coin tossing.
\newblock In {\em Proceedings of the International Conference on Computers,
  Systems and Signal Processing}, pages 175--179, 1984.

\bibitem{BBCM95}
Charles~H. Bennett, Gilles Brassard, Claude Crepeau, and Ueli~M. Maurer.
\newblock Generalized privacy amplification.
\newblock {\em IEEE Trans. Inf. Theor.}, 41(6):1915--1923, Nov 1995.

\bibitem{BBR88PA}
Charles~H. Bennett, Gilles Brassard, and Jean-Marc Robert.
\newblock Privacy amplification by public discussion.
\newblock {\em SIAM J. Comput.}, 17(2):210--229, Apr 1988.

\bibitem{BC90}
Samuel~L. Braunstein and Carlton~M. Caves.
\newblock {Wringing out better Bell inequalities}.
\newblock {\em Nuclear Physics B - Proceedings Supplements}, 6(0):211 -- 221,
  1989.

\bibitem{Cirelson1980}
Boris~S. Cirel'son.
\newblock Quantum generalizations of {B}ell's inequality.
\newblock {\em Letter in Mathematical Physics}, 4:93--100, 1980.

\bibitem{CHSH}
John~F. Clauser, Michael~A. Horne, Abner Shimony, and Richard~A. Holt.
\newblock Proposed experiment to test local hidden-variable theories.
\newblock {\em Phys. Rev. Lett.}, 23:880--884, Oct 1969.

\bibitem{ColRen11noextofQ}
Roger Colbeck and Renato Renner.
\newblock No extension of quantum theory can have improved predictive power.
\newblock {\em Nat. Commun.}, 2:411, Aug 2011.

\bibitem{Ekert91}
Artur~K. Ekert.
\newblock {Quantum cryptography based on Bell's theorem}.
\newblock {\em Phys. Rev. Lett.}, 67:661--663, Aug 1991.

\bibitem{HRWefficientDIQKD}
Esther H\"{a}nggi, Renato Renner, and Stefan Wolf.
\newblock Efficient device-independent quantum key distribution.
\newblock In {\em Proceedings of the 29th Annual International Conference on
  Theory and Applications of Cryptographic Techniques}, EUROCRYPT'10, pages
  216--234, 2010.

\bibitem{HRWimpossibilityPA}
Esther H\"{a}nggi, Renato Renner, and Stefan Wolf.
\newblock The impossibility of non-signaling privacy amplification.
\newblock {\em Theoretical Computer Science}, 486(0):27--42, 2013.

\bibitem{HILL99}
Johan Hastad, Russell Impagliazzo, Leonid~A. Levin, and Michael Luby.
\newblock A pseudorandom generator from any one-way function.
\newblock {\em SIAM J. Comput.}, 28(4):1364--1396, Mar 1999.

\bibitem{masanesPRL09}
Lluis Masanes.
\newblock Universally composable privacy amplification from causality
  constraints.
\newblock {\em Phys. Rev. Lett.}, 102:140501, Apr 2009.

\bibitem{MayYao98}
Dominic Mayers and Andrew Yao.
\newblock Quantum cryptography with imperfect apparatus.
\newblock In {\em Proceedings of the 39th Annual Symposium on Foundations of
  Computer Science}, FOCS '98, page 503, 1998.
%

\bibitem{PRbox}
Sandu Popescu and Daniel Rohrlich.
\newblock Nonlocality as an axiom.
\newblock {\em Foundations of Physics}, 24(379), (1994).

\bibitem{Ren08}
Renato Renner.
\newblock Security of quantum key distribution.
\newblock {\em International Journal of Quantum Information}, 6(01):1--127,
  2008.

\bibitem{EveNosignalling}
Benno Salwey.
\newblock {\em No-Signalling Attacks and Implications for (Quantum) Nonlocality
  Distillation}.
\newblock PhD thesis, USI Lugano, 2015.

\bibitem{VazVid14}
Umesh Vazirani and Thomas Vidick.
\newblock Fully device-independent quantum key distribution.
\newblock {\em Phys. Rev. Lett.}, 113:140501, Sep 2014.

\end{thebibliography}

\end{document}